\documentclass{aa}
\usepackage{graphicx, psfig, epsf, epsfig}

\newbox\grsign \setbox\grsign=\hbox{$>$}
\newdimen\grdimen \grdimen=\ht\grsign
\newbox\laxbox \newbox\gaxbox
\setbox\gaxbox=\hbox{\raise.5ex\hbox{$>$}\llap
	{\lower.5ex\hbox{$\sim$}}}\ht1=\grdimen\dp1=0pt
\setbox\laxbox=\hbox{\raise.5ex\hbox{$<$}\llap
	{\lower.5ex\hbox{$\sim$}}}\ht2=\grdimen\dp2=0pt

\newcommand{\xmm}{{\it{XMM-Newton}}}
\newcommand{\chandra}{{\it{Chandra}}}

\begin{document}
\title{Highly ionized Fe K$\alpha$ emission lines from the LINER galaxy M\,81
\thanks{Based on observations obtained with XMM-Newton, an ESA science mission
	  with instruments and contributions directly funded by ESA Member
	  States and the USA (NASA)}
}

   \author{
	  M.J. Page\inst{1}
	  R. Soria \inst{1},
	  S. Zane \inst{1},
	  K. Wu \inst{1} \and
	  R.L.C. Starling \inst{1} 
	  }
\authorrunning{M.J. Page et al.}
\titlerunning{Fe K$\alpha$ emission lines from M\,81}

\offprints{M.J. Page ({\tt mjp@mssl.ucl.ac.uk})}

\institute{$^{1}$ Mullard Space Science Laboratory, 
	  University College London, Holmbury St Mary, 
	  Surrey RH5 6NT, UK  \\
}

\date{Received 7 October 2003; accepted 8 April 2004}

\abstract{We present spectral and timing results from a long (130~ks) 
\xmm\ EPIC observation of the
nucleus of the Seyfert/LINER galaxy M\,81. During the observation the
X-ray flux varied by 20\%, but there was no significant change in spectral
shape. The 2--10 keV spectrum is well described by a power law continuum and
three narrow Fe~K$\alpha$ emission lines at 6.4, 6.7 and 6.96 keV. The three 
emission lines have equivalent widths of $39^{+13}_{-12}$, $47^{+13}_{-13}$ and
$37^{+15}_{-15}$~eV respectively. The ratios of the three lines are
thus more similar to those observed from the Galactic Centre region
than to those typically observed from Seyfert galaxies. The high ionization 
lines most likely
originate either from photoionized gas within 0.1 pc of the nucleus of M~81, or
from a non-thermal distribution of cosmic-ray electrons interacting with the
0.2-0.6 keV thermal plasma which is found in the bulge of M~81.
}

\authorrunning{M.J. Page et~al.}
   \maketitle
%
%________________________________________________________________

\section{Introduction}  

At a distance of 3.6 Mpc (Freedman et al. 1994), the Sab spiral M\,81 is the
closest galaxy showing the spectral signatures of a ``low ionization nuclear
emission-line region'' (LINER). LINER characteristics are found in a
significant fraction of all galaxies (between 1/5 and 1/3), and the nature of
the relationship between LINERs and low-luminosity AGN (LLAGN) has been a long
standing question (Ho, Filippenko \& Sargent \nocite{ho97} 1997, Heckman
\nocite{heckman80} 1980). The presence of a LLAGN in M\,81 is well established:
the unresolved nuclear source emits a non-stellar UV continuum, broad
optical/UV emission lines, and a significant flux of non-thermal X--rays (Ho,
Filippenko \& Sargent \nocite{ho96} 1996).  From dynamical studies, the mass of
the central compact object is found to be $\approx 4 \times 10^6\ {\rm
M}_{\odot}$ (Ho \nocite{ho99} 1999). With a bolometric luminosity of $\sim
10^{41}$~erg~s$^{-1}$, M\,81 is radiating at $< 10^{-3}$ of its Eddington
luminosity (Ho, Filippenko \& Sargent \nocite{ho96} 1996).

The nuclear X--ray source has a power law spectral shape in the 0.3-10 keV
energy range (Petre et al. \nocite{petre93} 1993, Ishisaki et
al. \nocite{ishisaki96} 1996, Swartz et al. \nocite{swartz03} 2003). The
spectrum is absorbed by a small column density ($\sim 3\times
10^{20}$~cm$^{-2}$) of material intrinsic to M\,81, and below 2 keV a multitude
of emission lines from circumnuclear, thermal gas can be seen in the \xmm\ RGS
spectrum of the nucleus (Page et al. \nocite{page03a} 2003).  Above 2 keV, the
most remarkable feature seen in the spectrum is an emission line at $6.7\pm0.1$
keV, corresponding to Fe~XXV~K$\alpha$ (Ishisaki et al. \nocite{ishisaki96}
1996, Pellegrini et al. \nocite{pellegrini00} 2000). Using data from {\em
BeppoSAX}, Pellegrini et al. \nocite{pellegrini00} (2000) claim that there is
an absorption edge at $\sim 8.6$ keV, which could also arise in Fe~XXV,
indicating that the nucleus is seen through a large column density ($\sim
2\times 10^{23}$~cm$^{-2}$) of highly ionized gas.  The 2-10 keV luminosity is
known to vary in the range $(1.5-6)\times 10^{40}$~erg~s$^{-1}$ on a timescale
of years, and variations of about 30\% have been observed on a timescale of
$10^{5}$~s (Pellegrini et al. \nocite{pellegrini00} 2000; Ishisaki et
al. \nocite{ishisaki96} 1996).

In this paper we examine the time variation and the 2-10 keV spectrum of the
M\,81 nucleus using the \xmm\ European Photon Imaging Camera (EPIC) data from
our long (130~ks) observation which was carried out as part of the RGS
consortium guaranteed time programme.

\section{Observations and data analysis}

M\,81 was observed with \xmm\ on 2001 April 22. The EPIC PN camera (Str\" uder
et al. \nocite{struder01} 2001) was operated in small window mode for a single
exposure of 130~ks duration.  Each of the two EPIC MOS cameras (Turner et
al. \nocite{turner01} 2001) were exposed for 122~ks, split into two
exposures. MOS1 was operated in timing mode, while MOS2 was operated in
full-frame imaging mode. All three EPIC cameras were used in conjunction with
medium filters.  The nuclear X--ray source suffers from significant photon
pile-up in the MOS2 data, and so these data are not included in our
analysis. Pile-up is negligible for the nuclear source in the MOS1 and PN
exposures.

The MOS1 data were processed with SAS v4.1. For technical reasons, it was not
possible to reprocess the PN data using SAS v4.1, so they were processed using
SAS v5.3.3.  In order to construct lightcurves and spectra, source counts were
extracted from a 44\arcsec\ radius circular region in the PN camera, and from a
39\arcsec\ wide strip in the MOS1 camera.  Both regions were centred on the
nucleus and correspond to an encircled energy fraction of $88\%$ of the nuclear
flux. In high resolution {\em Chandra} imaging of M\,81, the total X-ray flux
originating from diffuse emission and point sources between 10\arcsec\ and
44\arcsec\ from the nucleus is only $\sim 2\%$ of the nuclear flux (Swartz
et~al. \nocite{swartz03} 2003).  Hence the EPIC spectra are completely
dominated by emission arising from the nucleus and its immediate environment
(within $\sim 200$pc).  In both cameras, background counts were taken from
regions which were free from bright sources. In the PN camera these were on the
same chip as the source region, and in the MOS1 camera they were in the inner
parts of the outer chips.

Lightcurves were constructed from both PN and MOS 1 data in the energy bands
0.3-2.0~keV and 2-10~keV with 1000\,s time bins. To maximise the signal to
noise, the lightcurves from the separate instruments were added to produce a
single lightcurve for each energy band.

When extracting spectra, we excluded periods of high background flux. This
reduces the durations of the exposures to 119~ks for the PN camera and 93~ks
for MOS1. Response matrices and effective area files specific to the source
regions were constructed using the standard {\tt RMFGEN} and {\tt ARFGEN} SAS
tasks. To optimise signal to noise the spectra from MOS1 and PN were combined
using the method of Page, Davis and Salvi \nocite{page03b} (2003) in 45~eV bins
which sample the EPIC spectral resolution well. The resultant spectrum has
$\sim 400$ counts per bin at 6.4 keV and $> 40$ counts in every bin from 2 to
10 keV.  Spectral analysis was carried out with {\tt XSPEC} 11.1.0 (Arnaud
\nocite{arnaud96} 1996).

\section{Results}

\subsection{Time variability}
\label{sec:lightcurves}

\begin{figure} 
\vspace*{0.25cm} 
\psfig{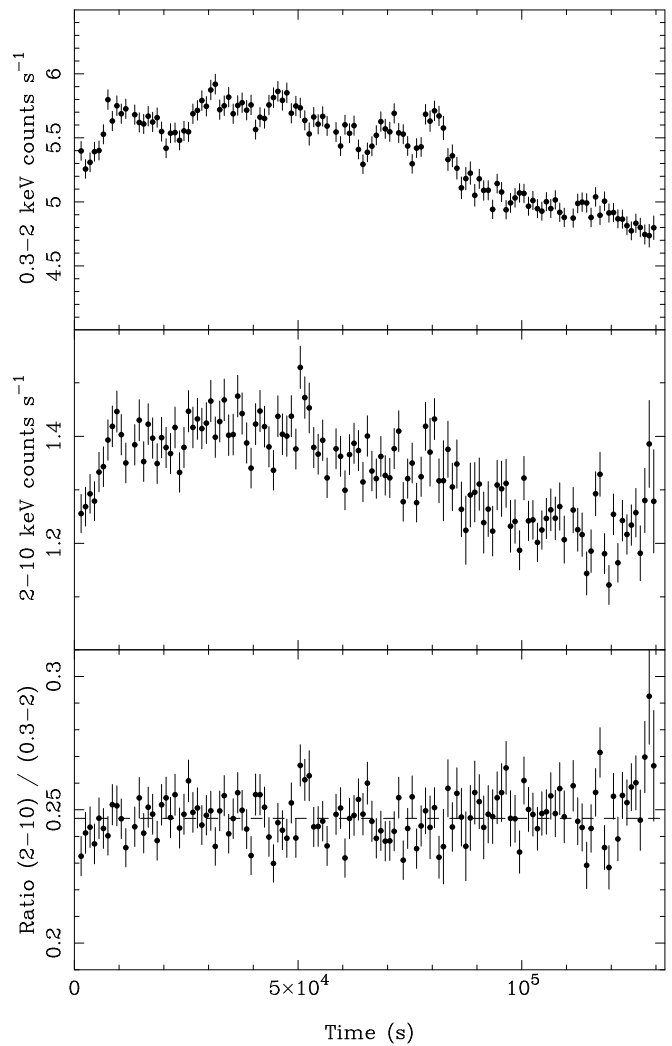}
\caption{EPIC lightcurves with 1000\,s time bins in the 0.3-2.0 keV band (top
panel) and 2-10 keV band (middle panel). The bottom panel shows the ratio of
the two lightcurves; the dashed line shows a constant fit to the hardness ratio
(see Section \ref{sec:lightcurves}).}
\label{fig:lightcurves}
\end{figure}

The background-subtracted EPIC PN+MOS1 lightcurves for the 0.3-2.0~keV and
2-10~keV bands are shown in the top and middle panels of
Fig.~\ref{fig:lightcurves} respectively. The lightcurves show flickering,
similar in character to the stochastic variability commonly seen in Seyfert
galaxies (e.g. Green, McHardy \& Lehto \nocite{green93} 1993), but of
relatively low amplitude ($\sim 20\%$ peak to trough).  The ratio of the
2-10~keV countrate to the 0.3-2 keV countrate is shown in the bottom panel of
Fig.\ref{fig:lightcurves}. A constant fit to these data points (shown as dashed
line in Fig.\ref{fig:lightcurves}) results in $\chi^{2}/\nu = 141/125$,
corresponding to a null hypothesis probability of 15\%. This means that the low
level variability is not accompanied by any significant gross spectral changes.

\subsection{Spectral study}    
\label{sec:spectrum}

\begin{figure} 
\begin{center}
\psfig{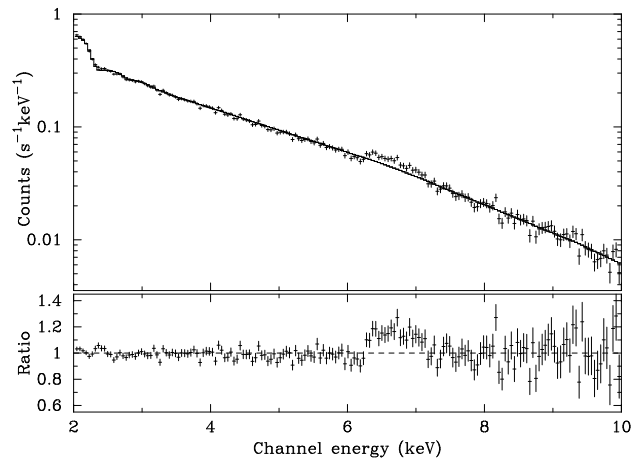}
\caption{Top panel: EPIC spectrum of M\,81 (datapoints) and the best fit power
law model (stepped line). Bottom panel: ratio of the datapoints to the power
law model.}
\label{fig:powerlaw}
\end{center}
\end{figure}

\begin{figure} 
\begin{center}
\psfig{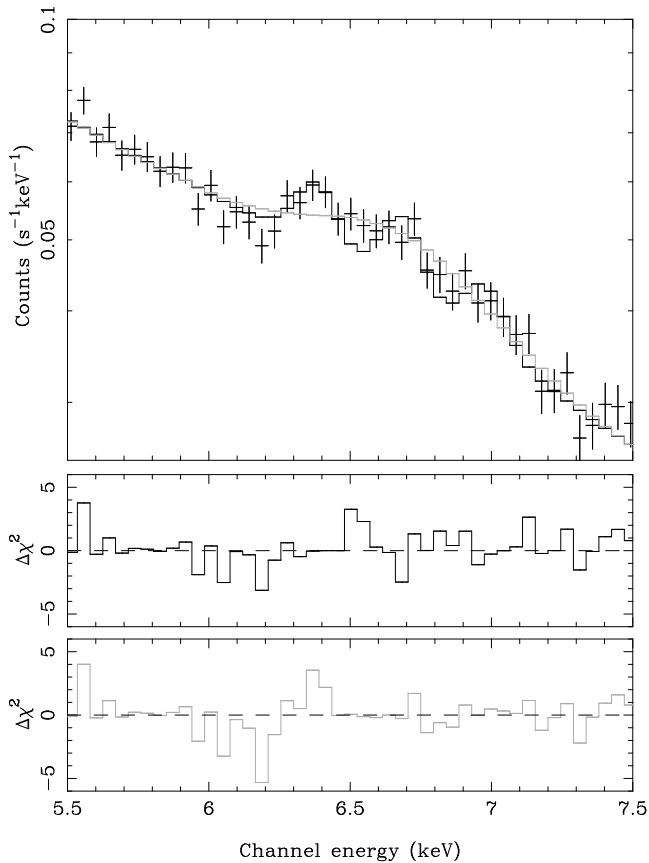}
\caption{Top panel: EPIC spectrum of M\,81 (datapoints), the best fit power law
+ three Fe emission lines model (black stepped line), and the power law + broad
Gaussian model (grey stepped line). Middle panel: contribution of the
datapoints to the $\chi^2$ multiplied by the sign (model-data) for the power
law + three Fe emission lines model. Bottom panel: contribution of the
datapoints to the $\chi^2$ multiplied by the sign (model-data) for the power
law + broad Gaussian model.}
\label{fig:felines}
\end{center}
\end{figure}

\begin{figure} 
\begin{center}
\psfig{figure=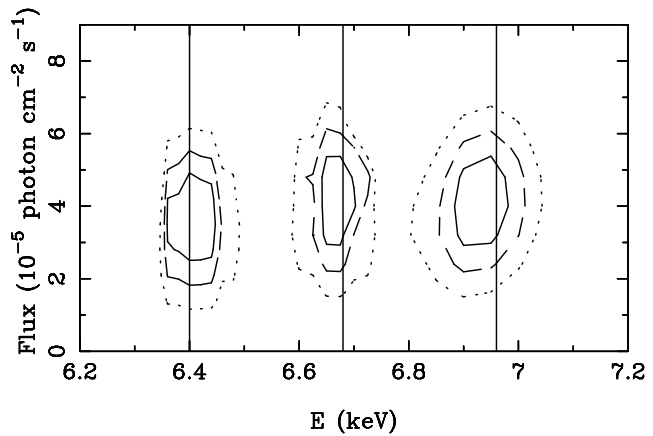,width=8.5cm}
\caption{Confidence intervals for the energies and normalisations of the three
narrow emission lines used to fit the bump between 6.25 and 7.15 keV. The
solid, dashed and dotted contours correspond to 1, 2 and 3$\sigma$ confidence
for two interesting parameters respectively. The solid vertical lines indicate
the expected energy for Fe~I, Fe~XXV and Fe~XXVI.}
\label{fig:contours}
\end{center}
\end{figure}

The 2-10~keV EPIC spectrum of M\,81 is shown in Fig. \ref{fig:powerlaw}.  We
began by fitting a power law model, absorbed by neutral material in our own
Galaxy and within M\,81, with the total column density fixed at the value of
$7.6\times 10^{20}$~cm$^{-2}$, as determined from the simultaneous RGS
observation (Page et al. \nocite{page03a} 2003). This model, and the data/model
ratio are shown in Fig.  \ref{fig:powerlaw}. The parameters for this fit and
all subsequent fits are given in Table \ref{tab:fits}.  While the overall
continuum shape is well reproduced by the power law, the $\chi^{2}$ is very
poor ($\chi^{2}/\nu=323/175$), and the data show a significant excess over the
model between 6.25 and 7.15 keV.  The energy range of the excess emission
corresponds to the K$\alpha$ lines of Fe~I -- Fe~XXVI, although a photoelectric
edge from Fe~I at 7.1~keV could also contribute to the excess if there is
significant reflection from cold material. Therefore we added a narrow 6.40~keV
Gaussian emission line corresponding to Fe~I K$\alpha$ and a neutral Compton
reflection component (Magdziarz \& Zdziarski \nocite{magdziarz95} 1995; {\tt
PEXRAV} in {\tt XSPEC}) to the absorbed power law model. Both these components
could arise in a cold reflecting medium, e.g. an accretion disc or a distant
molecular torus (George \& Fabian \nocite{george91} 1991; Matt, Perola \& Piro
\nocite{matt91} 1991); the inclination of the reflector was fixed at 60~deg to
the observer's line of sight in the fit. The fit is much improved
($\Delta\chi^{2}$ of 69 for 2 extra free parameters), compared to the absorbed
power law model. However, this model is still rejected with $>99.9\%$
confidence, and the data showed a significant excess over the model at
$\sim6.7$~keV. Therefore we added a second narrow Gaussian emission line, at
6.68~keV, corresponding to Fe~XXV. The fit is improved significantly with the
addition of the second emission line ($\Delta\chi^{2}=30$ for 1 extra free
parameter).  Although the model is now not rejected at the the 99\% confidence
level, it is still rejected at the 95\% level, and the data still show an
excess over the model between 6.8 and 7.15 keV.  Therefore we included a third
narrow Gaussian emission line at 6.96~keV, corresponding to Fe~XXVI, in the
model. This results in another significant improvement in the fit
($\Delta\chi^{2}=17$ for 1 extra free parameter) and the fit is now acceptable
at the 95\% confidence level. However, as shown in Table \ref{tab:fits}, when
the three Fe~K$\alpha$ emission lines are included in the fit, the amplitude of
the reflection component is consistent with zero. We therefore removed the
reflection component from the model and repeated the fit. The $\chi^{2}$ is the
same for this model fit as it is when the reflection component is included, but
with one fewer free parameter.  Hence our best fit model is a power law with
three Gaussian emission lines corresponding to Fe~I, Fe~XXV and Fe~XXVI; this
model is shown along with the data in Fig. \ref{fig:felines}. The 2-10 keV flux
inferred from the fit is 1.1~$\times$~$10^{-11}$~erg~s$^{-1}$~cm$^{-2}$.  To
further check the consistency of the line associations with Fe~I, Fe~XXV and
Fe~XXVI we repeated the fit allowing the energies of the emission lines to
vary. This results in a negligible improvement in goodness of fit
($\Delta\chi^{2}=1$ for 3 extra parameters) and the emission line parameters
are all consistent with those obtained when the line energies were fixed (see
Fig. \ref{fig:contours}).

An alternative interpretation for the 6.25-7.15 keV emission is that it is
produced by a single broad emission line rather than by a combination of narrow
lines. When a single broad Gaussian is tried, we obtain formally a good fit,
with $\chi^{2}/\nu=204/172$. However, this model does not reproduce the sharp
6.25~keV shoulder of the emission feature, and is therefore less satisfactory
than the multi-line fit (see Fig. \ref{fig:felines}).  The best fit central
energy of the line is $6.69^{+0.06}_{-0.07}$ keV consistent with
Fe~XXV~K$\alpha$, and the full width half maximum (FWHM) of the line is
$28000\pm 7000$~km~s$^{-1}$.

\begin{figure} 
\begin{center}
\psfig{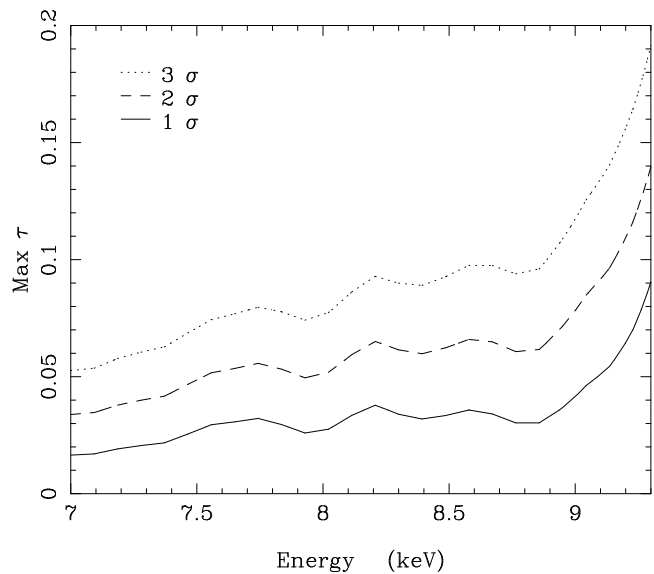}
\caption{Confidence contour (essentially upper limits) for a photoelectric edge
due to ionized Fe. The solid, dashed, and dotted lines correspond to confidence
intervals of 1, 2 and 3$\sigma$ respectively for two interesting parameters.}
\label{fig:edgecont}
\end{center}
\end{figure}

Pellegrini et al. \nocite{pellegrini00} (2000) claimed to have detected a
photoelectric edge at 8.6$^{+0.4}_{-0.8}$ keV with optical depth
$\tau=0.15^{+0.09}_{-0.08}$ in the spectrum of M\,81, which they hypothesized
could be due to Fe~XXV.  We have searched for Fe~K photoelectric edges in our
spectrum by adding an edge to our best fit model. The threshold energy of the
edge was allowed to vary between 7 and 9.3 keV, corresponding to K edges from
Fe~I -- Fe~XXVI. However, we find that the $\chi^{2}$ is unchanged by this
extra model component, and the best fit optical depth for the edge is zero,
regardless of its energy. To determine upper limits we stepped the edge energy
and optical depth over a grid of values, to produce the confidence contour
shown in Fig. \ref{fig:edgecont}.

\begin{table*}
\begin{center}
\caption{Results of spectral fits to the EPIC data. All fits were performed
over the 2-10 keV energy range. All the spectral models included absorption
from neutral material with a column density of $7.6\times 10^{20}$~cm$^{-2}$.
Uncertainties are quoted at 95\% for 1 interesting parameter
(i.e. $\Delta\chi^{2}=4$). $A_{P.L.}$ is the power law normalisation in units
of $10^{-3}$~photons~cm$^{-2}$~s$^{-1}$~keV$^{-1}$.  $R$ is the solid angle of
the sky covered by the reflector as seen by the X--ray source in units of
$2\pi$ steradians. When a parameter was constrained by the limits of the
fit-range rather than by $\Delta \chi^{2}$ the corresponding limit has been
marked with an `*'.}
\label{tab:fits}
\begin{tabular}{lcccccccc}
&&&6.40 keV&6.68 keV&6.96 keV&&&\\
Model& $\Gamma$ & $A_{P.L.}$ & E.W. & E.W. & E.W.
& $R$ & $\chi^{2}/\nu$ & Prob\\
&&&(eV)&(eV)&(eV)&&&\\
&&&&&&&&\\
P.L.&$1.86^{+0.01}_{-0.02}$&$3.55^{+0.07}_{-0.07}$
&-&-&-&-&323/175&$7.4\times 10^{-11}$\\
P.L.+Gau+refl&$1.95^{+0.05}_{-0.04}$&$3.87^{+0.18}_{-0.14}$&
$26^{+13}_{-13}$&-&-&$1.9^{+0.9}_{-0.6}$&245/173&$2.5\times 10^{-4}$\\
P.L.+2$\times$Gau+refl&$1.92^{+0.05}_{-0.05}$&$3.78^{+0.15}_{-0.13}$&
$30^{+13}_{-13}$&$38^{+15}_{-14}$&-&$1.2^{+0.8}_{-0.6}$&218/172&
$1.0\times 10^{-2}$\\
P.L.+3$\times$Gau+refl&$1.88^{+0.05}_{-0.05}$&$3.68^{+0.17}_{-0.15}$&
$35^{+13}_{-13}$&$43^{+15}_{-15}$&$32^{+16}_{-16}$&$0.6^{+0.8}_{-0.6*}$
&202/171&
$5.3\times 10^{-2}$\\
P.L.+3$\times$Gau&$1.88^{+0.02}_{-0.02}$&$3.64^{+0.07}_{-0.07}$&
$39^{+13}_{-12}$&$47^{+13}_{-13}$&$37^{+15}_{-15}$&-&202/172&
$5.8\times 10^{-2}$\\
P.L.+broad Gau$^{\dagger}$&$1.88^{+0.02}_{-0.01}$&$3.65^{+0.07}_{-0.07}$&
-&$160^{+40}_{-40}$&-&-&204/172&
$4.8\times 10^{-2}$\\
&&&&&&&&\\
\multicolumn{9}{l}{$\dagger$ In this model the central energy and width 
$\sigma$ of the emission
line are free parameters. The best fit values are $6.69_{-0.07}^{+0.06}$~keV}\\
\multicolumn{9}{l}{and $0.27_{-0.06}^{+0.07}$~keV respectively.}\\
\end{tabular}
\end{center}
\end{table*}

\section{Discussion}  

In common with previous investigations, we find that the 2-10 keV spectrum is
dominated by a power law continuum.  The spectral slope we obtain
($\Gamma=1.88\pm0.02$) is consistent with that seen below 2 keV in the
simultaneous RGS observation ($\Gamma=1.94\pm0.06$; Page et
al. \nocite{page03a} 2003). Low level variability, of around 20\% peak to
trough is observed during the observation, but there are no detectable spectral
changes.  Similar fractional variations in intensity, over 1--2 day timescales,
have also been observed in {\it{ASCA}} (Ishisaki et al. \nocite{ishisaki96}
1996) and {\it{BeppoSAX}} (Pellegrini et al.~2000) observations.  The mean EPIC
countrate during the \xmm\ observation corresponds to a 2-10 keV flux of
1.1~$\times$~$10^{-11}$~erg~s$^{-1}$~cm$^{-2}$, around 30\% of the flux at the
time of the {\it{BeppoSAX}} observation. Thus the X--ray variability of M\,81
observed on a 1--2 day timescale is consistent with the linear relationship
between flux and variability amplitude that is observed in Seyfert galaxies,
whereby the variability amplitude is proportional to the mean flux, and so the
fractional variations change little from observation to observation even though
the mean flux may change considerably (Uttley \& McHardy \nocite{uttley01}
2001). On the other hand, the variability amplitude of M\,81 (and LINERs in
general) is much lower than would be expected from extrapolating Seyfert
variability amplitudes to lower luminosity (Ptak et~al.  \nocite{ptak98} 1998).
The spectral index obtained from the \xmm\ observation is consistent with those
obtained from observations with {\it ASCA} ($\Gamma=1.85\pm0.04$, Ishisaki et
al. \nocite{ishisaki96} 1996) and with {\it{BeppoSAX}} ($\Gamma=1.86\pm0.03$,
Pellegrini et al. \nocite{pellegrini00} 2000).  Thus the continuum shape
appears to be unaffected by intensity variations either over timescales of days
or over timescales of years.

\subsection{Emission and absorption from ionized Fe}

The high quality of our EPIC spectrum allows us to resolve the Fe~K$\alpha$
line emission, which can be modelled either by a single broad Fe~XXV emission
line or as the sum of three narrow components: Fe~XXVI, Fe~XXV, and a component
representing emission from lower ionisation Fe species.  Fe~K$\alpha$ emission
is commonly seen in Seyfert galaxies, and is thought to be due to reprocessing
of the primary X--rays by surrounding material.  If the material lies along the
line of sight, we will see both absorption and emission features in the X--ray
spectrum. This is the hypothesis favoured by Pellegrini et
al. \nocite{pellegrini00} (2000), because their {\it{BeppoSAX}} spectrum
appeared to show both an emission line and an absorption edge from Fe~XXV.
However, we can rule out a Fe~XXV K edge with significant optical-depth ($\tau
> 0.1$) at the $>99.7\%$ confidence level in our EPIC spectrum.  One possible
explanation for this discrepancy might be that the X--ray flux at the time of
the EPIC observation was only about 30\% of the flux at the time of the
{\it{BeppoSAX}} observation: assuming the absorber is photoionized, the
corresponding reduction in the ionization level of the absorber might explain
the lack of an Fe~XXV edge in the EPIC spectrum.  If this were the case, we
would expect the dominant ionic species of Fe in the absorber to change from
Fe~XXV during the {\it{BeppoSAX}} observation to Fe~XX -- Fe~XXIV in the \xmm\
observation (Kallman \& McCray \nocite{kallman82} 1982).  Instead of the $\tau
\sim 0.2$ Fe~XXV K edge at 8.8 keV, there should be 1-4 Fe~K edges between 8.0
and 8.6 keV. The relative ionic abundances of Fe at this ionization level
(Kallman \& McCray \nocite{kallman82} 1982), and the corresponding K-shell
photoionization cross sections (Verner \& Yakovlev \nocite{verner95} 1995),
imply that at least one of these should have $\tau > 0.06$ at threshold, but
this is ruled out with 95\% confidence by our upper limits given in
Fig.~\ref{fig:edgecont}.

An absorber in which the dominant species of Fe are Fe~XX -- Fe~XXIV should
also produce a host of Fe~L absorption lines in the soft X--ray. The deepest of
these lines lie at 12.82\,\AA, 12.29\,\AA, 11.71\,\AA, 11.00\,\AA, and
10.64\,\AA\ for Fe~XX -- Fe~XXIV respectively (e.g. Verner, Verner \& Ferland
\nocite{verner96} 1996). If the column density implied by the Fe~XXV K edge
found by Pellegrini et al. \nocite{pellegrini00} (2000) were shared evenly
between these ions, and assuming a velocity dispersion of $\sigma >
50$~km~s$^{-1}$ in the absorber, we would expect these lines to have equivalent
widths of at least 29\,m\AA, 27\,m\AA, 19\,m\AA, 15\,m\AA\ and 13\,m\AA\
respectively.  We have looked for these absorption lines in the simultaneous
RGS spectrum of M~81 studied by Page et al. \nocite{page03a} (2003) by
repeating their spectral fit, adding these minimum absorption lines one at a
time to their best fit model, and measuring the $\Delta \chi^{2}$. For every
ion except Fe~XXII the absorption line makes the $\chi^{2}$ poorer, by $\Delta
\chi^{2}$ values in the range 1.4 to 9. For Fe~XXII, adding an absorption line
of 19\,m\AA\ equivalent width improves the $\chi^{2}$ by 0.9. Increasing the
equivalent widths of any of these lines (including Fe~XXII) from their expected
minima results in an increased $\chi^{2}$.  Furthermore, in this range of
ionization, 3--20\% of O should be in the form of O~VIII. Assuming the absorber
has solar abundances (Anders \& Grevesse \nocite{anders89} 1989), and a
velocity distribution with $\sigma > 50$~km~s$^{-1}$, the absorber should give
rise to O~VIII Ly$\alpha$ and Ly$\beta$ absorption lines with equivalent widths
of $>28$\,m\AA\ and $>16$\,m\AA\ respectively. Such absorption lines are not
seen in the simultaneous RGS spectrum (Page et al. \nocite{page03a}
2003). Instead, O~VIII Ly$\alpha$ is seen in emission, and while this might
mask the absorption line, the corresponding Ly$\beta$ emission is much weaker.
To see whether such a Ly$\beta$ absorption line might be present in the RGS
spectrum, we have again repeated the spectral fit of Page et
al. \nocite{page03a} (2003) to the RGS spectrum, adding an absorption line at
the position of O~VIII Ly$\beta$ to their best fit model. The best fit
equivalent width is zero, and an absorption line of $16$\,m\AA\ equivalent
width results in a $\Delta \chi^{2}$ of 4, and so is excluded at 95\%
confidence.  In fact we do not see any evidence for ionized absorption in the
RGS spectrum.  Therefore we conclude that the lack of an Fe~XXV K absorption
edge in our EPIC spectrum is not explained by the reduction of the
ionization. Either the ionized absorber has moved out of the line of sight, or
the $2.8\sigma$ edge feature in the {\it{BeppoSAX}} spectrum was a statistical
fluctuation rather than a real feature.  Therefore, if the Fe~K$\alpha$ line(s)
are reprocessed emission from the X--ray radiation, they come from outside the
line of sight.

If a broad Fe~XXV line is responsible for all the emission, the large velocity
width (FWHM~=~$28000\pm7000$~km~s$^{-1}$) implies that the emission originates
within a few hundred Schwarzschild radii of the compact object. Furthermore,
because the emission feature has a relatively high equivalent width
($160\pm40$~eV), it must have a significant Thomson optical depth (i.e. $>0.1$)
and cover a significant fraction of the sky as seen by the X-ray source (see
Bianchi \& Matt \nocite{bianchi02} 2002).  The inner part of an accretion disc
is the only viable location for gas with this combination of properties, and
reflection from an accretion disc surface is widely hypothesized for broad
Fe~K$\alpha$ emission in AGN.  However, the inner part of an accretion disc is
unlikely to be responsible for broad 6.7 keV line emission in M~81, on both
theoretical and empirical grounds.  Firstly, because of its low bolometric
luminosity ($L<10^{-3}L_{\rm Edd}$), M~81 probably contains an advective flow
in the central regions rather than an accretion disc (Narayan \& Yi
\nocite{narayan95} 1995). This hypothesis is further supported by the lack of
any soft, blackbody-like component in the soft X--ray spectrum of M~81 (Page
et~al. \nocite{page03a} 2003); such a component is commonly observed in Seyfert
galaxies and is thought to be the high energy tail of the emission from the
inner accretion disc (Arnaud et~al. \nocite{arnaud85} 1985).  The second reason
that a broad Fe~XXV line is unlikely to originate in an accretion disc in M\,81
is that the X--ray spectrum lacks any of the other features expected from such
a disc. In particular, the soft X-ray spectrum does not contain the strong,
broad emission features predicted to occur in discs which are sufficiently
ionized to produce significant Fe~XXV emission (e.g. O~VIII and O~VII emission
lines and radiative recombination continua; see Nayakshin \& Kallman
\nocite{nayakshin01} 2001 and Ballantyne, Ross \& Fabian \nocite{ballantyne01}
2001).  Instead, the only emission features observed in the soft X-ray spectrum
are the narrow emission lines from a spatially extended thermal component (Page
et~al. \nocite{page03a} 2003). Thus a single broad Fe~XXV emission line, as
well as failing to reproduce the sharp 6.25~keV shoulder of the emission
feature, is physically inconsistent with the other X-ray and multiwavelength
properties of M~81.  However, for a novel explanation for a broad 6.7 keV
emission line, arising at the interface between the thin, outer disc and the
inner, radiatively-inefficient, hot flow, see Dewangan
et~al. \nocite{dewangan04} (2004).

Because the broad line model for the Fe~K$\alpha$ emission appears unlikely on
physical grounds, we consider the combination of narrow 6.4, 6.7 and 6.9 keV
emission lines.  The three components each contribute about a third of the line
emission, and the combined equivalent width is $120\pm20$~eV.  For the 6.4 keV
emission line, which arises in cool material, the observed equivalent width of
$\sim 40$~eV is rather low compared to the line width observed in most Seyfert
galaxies (typically $\sim 100$~eV, e.g. Nandra et~al. \nocite{nandra97}
1997). If the line arises in a Compton thick, plane parallel reflector, then
the small equivalent width implies either that the reflector is highly inclined
to our line of sight (i.e. $i > 80^{o}$) or that the reflector covers less than
50\% of the sky as seen by the X-ray source (George \& Fabian \nocite{george91}
1991). If the reflector has a funnel shape (e.g. the molecular torus of AGN
unification schemes) then the low equivalent width implies an even lower
sky-coverage of the reflector (Matt et~al.  \nocite{matt91}
1991). Alternatively the low equivalent width of the 6.4 keV line could be
explained if the reprocessing region is Compton-thin (Leahy \& Creighton
\nocite{leahy93} 1993).  In any case, the low equivalent width of the 6.4 keV
line implies that in M\,81 there is not as much cold, X-ray illuminated
material as there is in a typical Seyfert galaxy.  In contrast, the 6.7 and
6.96 keV lines have a larger equivalent width in M\,81 than is normally
observed in Seyfert~1 galaxies, which typically have 6.4 keV lines that are at
least four times stronger than either the 6.7 or 6.96 keV line; indeed the 6.7
or 6.96 keV lines are often not detectable in EPIC X-ray spectra of Seyfert~1s
(e.g. Pounds et~al. \nocite{pounds03} 2003, Blustin et~al. \nocite{blustin03}
2003).  The calculations by Bianchi \& Matt \nocite{bianchi02} (2002) allow us
to place some interesting constraints on photoionized material that might give
rise to these lines. According to these calculations the column density of
illuminated, reflecting material must be $\ge 10^{23}$~cm$^{-2}$ in order to
produce the observed equivalent widths. To maintain a sufficiently high level
of ionization in the illuminated material to produce the 6.7 and 6.96 keV lines
requires that the material lies quite close to the illuminating source.  From
equation 6 of Bianchi \& Matt \nocite{bianchi02} (2002), and for the observed
2-10 keV luminosity of M\,81 ($1.7\times 10^{40}$~erg~s$^{-1}$), we find that
the illuminated material responsible for the 6.7 keV line must lie within
0.1~pc of the nucleus, while the material responsible for the 6.96 keV line
must lie within 0.01~pc.

Although the 6.4-7 keV complex could not be separated into its components in
the {\it ASCA} and {\it{BeppoSAX}} observations, the total equivalent width
measured with these data was consistent with our EPIC value, at $170\pm60$~eV
(Ishisaki et al. \nocite{ishisaki96} 1996) and $95^{+28}_{-46}$~eV (Pellegrini
et al. \nocite{pellegrini00} 2000) respectively. Unfortunately, the
uncertainties on these measurements are large enough that they are also
consistent with constant flux from the lines (i.e. a reduction in the
equivalent width by a factor of three between the {\it{BeppoSAX}} and \xmm{}
observations).

\subsection{Comparison to the Galactic Centre}

It is interesting to note that the Fe~K$\alpha$ line ratios are relatively
similar to those observed in the 1~deg~$\times$~1~deg (150~pc $\times$ 150~pc)
region around our own Galactic Centre (Tanaka et al. \nocite{tanaka00} 2000).
In particular, emission lines at 6.4 keV, 6.7 keV and 6.9 keV are seen in both
the Galactic Centre and in M\,81, and in both cases the 6.7 keV line is the
strongest (Table \ref{tab:fits} and Tanaka \nocite{tanaka02} 2002).  In the
case of the Galactic Centre, the origin of the Fe~K lines is controversial
(e.g. Ebisawa et al. \nocite{ebisawa03} 2003; Predehl et al. \nocite{predehl03}
2003).  The 6.4 keV emission is thought to arise through fluorescence in cold
Fe, which is either illuminated by X--rays emitted during a past, comparatively
more active period of Sgr~A$^{*}$ (e.g. Murakami, Koyama \& Maeda
\nocite{murakami01} 2001), or bombarded by low energy cosmic ray electrons
(Valinia et al. \nocite{valinia00} 2000).

Possible models for the 6.7~keV and 6.96~keV lines include a high temperature
($kT\sim$10 keV) thermal plasma (Koyama et al. \nocite{koyama89} 1989), charge
exchange between cosmic-ray Fe ions and interstellar H (Tanaka, Miyaji \&
Hasinger \nocite{tanaka99} 1999), interaction of non-thermal electrons with a
low temperature ($\sim 0.3$~keV) thermal plasma (Masai et al. \nocite{masai02}
2002; Dogiel et al. \nocite{dogiel02} 2002; Valinia et al. \nocite{valinia00}
2000), or a large population of faint accreting binaries (Wang et
al. \nocite{wang02} 2002).  However, there now appear to be significant
problems with the first two of these models for the 6.7 and 6.96 keV lines: a
10 keV thermal plasma is too hot to be gravitationally bound to the Galaxy, and
so strong magnetic fields ($\sim 30 \mu$G) would be required to confine the
plasma (Tanuma et al. \nocite{tanuma99} 1999), while charge exchange between
cosmic ray Fe ions and interstellar H is expected to produce 8.5 and 8.9 keV
lines rather than the 6.7 and 6.96 lines which are observed (Masai et
al. \nocite{masai02} 2002).

Although the integrated contribution of accreting binary stars may plausibly
explain the 6.7 and 6.96 keV lines from the Galactic Centre region (Wang et~al
\nocite{wang02} 2002), this would seem unlikely for M\,81.  The population of
discrete X--ray sources near the Galactic Centre has a radial surface density
which declines as $R^{-1}$ where $R$ is distance from the centre (Muno
et~al. \nocite{muno03} 2003).  This is similar to the spatial distribution of
stars selected in the infrared (Serabyn \& Morris \nocite{serabyn96} 1996),
suggesting that the X-ray source population is intimately linked to the
underlying population of bulge stars.  Furthermore, the luminosity function of
the discrete sources is steep, with $dN/dL \propto L^{-1.7}$, at the faintest
fluxes probed by Muno et~al. \nocite{muno03} (2003) so that the emission from
discrete sources is dominated by low luminosity objects (2--8 keV luminosity
$<10^{32}$~erg~s$^{-1}$), likely to be low mass accreting binaries such as
cataclysmic variables. Such objects show strong Fe~XXV and Fe~XXVI K $\alpha$
lines (see e.g. Mukai et~al. \nocite{mukai03} 2003; Wu, Cropper \& Ramsay
\nocite{wu01} 2001) and will predominantly be drawn from the old stellar bulge
population because they are long lived (Warner \nocite{warner95} 1995).  The
bulges of the Milky Way and M\,81 have similar overall luminosities: $5 \times
10^{9}$ L$_{\odot}$ for M\,81 (determined from V-band photometry: Tenjes, Haud
\& Einasto \nocite{tenjes98} 1998), and $6-11 \times 10^{9}$ L$_{\odot}$ for
the Milky Way (determined from near-IR photometry: Kent, Dame \& Fazio
\nocite{kent91} 1991; Freudenreich \nocite{freudenreich98} 1998).  We would
therefore expect similar integrated Fe~K line luminosities from the low
luminosity binary populations in the central regions of M\,81 and the Milky
Way.  The luminosity of the Fe~K$\alpha$ lines emitted near the Galactic Centre
is $1.1\pm0.1 \times 10^{36}$~erg~s$^{-1}$ (Yamauchi et al. \nocite{yamauchi90}
1990), and assuming a distance of 3.6 Mpc to M\,81, the Fe~K$\alpha$ line
emission from M\,81 is $1.6\pm0.3 \times 10^{38}$~erg~s$^{-1}$, about 150 times
more luminous than that from the Galactic Centre.  Therefore, whatever fraction
of the Fe~K line emission in the Galactic Centre comes from accreting binaries,
such sources are unlikely to make a substantial contribution to the Fe~K lines
in the central part of M\,81.

On the other hand, if the Galactic Centre Fe lines originate from a non-thermal
distribution of cosmic-ray electrons, interacting with both cold and hot
(0.2-0.6 keV) gas, the Fe lines in M\,81 could quite plausibly arise through
the same mechanism. In the radio band it has already been noted that the
Galactic Centre and M\,81 appear to be remarkably similar sources, with M\,81
resembling a scaled-up version of the Galactic Centre, from the core-jet
structures (M\,81* and Sgr~A*) seen on sub-pc scales (Bietenholz, Bartel \&
Rupen \nocite{bietenholz00} 2000; Lo et~al. \nocite{lo98} 1998) to the large
arc-like structures seen on tens to hundreds of pc scales (Kaufman
et~al. \nocite{kaufman96} 1996; Yusef-Zadeh, Morris \& Chance \nocite{yusef84}
1984). The radio luminosity of M\,81* is about $10^{4}$ times higher than that
of Sgr~A*, which suggests that the non-thermal electron population in the
central regions of M\,81 could be supplied with sufficient energy to power the
Fe~K$\alpha$ emission.  Furthermore, myriad soft X-ray emission lines from
0.2-0.6 keV thermal plasma have already been detected from the central kpc of
M\,81 where the most important radio structures are found.

The spatial extent of the 6.4, 6.7 and 6.9 keV line emission in M\,81 will be a
key diagnostic to determine whether they are analogous to those observed from
the Galactic Centre or are due to photoionization by the nuclear X--ray source.
The lines in the Galactic Centre come from a region more than 100~pc across; if
the M\,81 lines are a scaled-up version of the same phenomenon, then they can
be expected to arise from region of similar size or larger (i.e. extending out
to at least 6 arcseconds from the nucleus of M\,81).  On the other hand, if the
lines are photoionized by the nucleus they must originate within 0.1~pc of the
nucleus (6 milli-arcseconds at the distance of M\,81). Unfortunately, at the
spectral resolution of the EPIC CCDs the peak intensity of the lines is at most
$\sim 20\%$ of the underlying AGN continuum, and \xmm's spatial resolution is
not sufficiently good that we can distinguish such a weak component of extended
emission. \chandra\ has much better spatial resolution, but it has a much lower
sensitivity than EPIC around Fe~K, and the existing observations suffer badly
from photon pile-up close to the nuclear source.  Currently, Fe~K emission has
been detected neither from the nuclear source, nor from the circumnuclear 10-30
arcsecond region (Swartz et~al. \nocite{swartz03} 2003). In principle, a very
long observation using the high-energy transmission gratings on \chandra\ might
allow the measurement of the spatial extent of the Fe~K line emitting region in
M\,81 because it could achieve the required spatial {\em and} spectral
resolution without significant photon pile-up.

\section{Conclusions}

The 2-10 keV X-ray spectrum of the nucleus of M\,81 consists of a power law
continuum and three Fe~K$\alpha$ emission lines at 6.4, 6.7 and 6.96 keV.  The
power law component has a photon index $\Gamma \sim 1.9$, consistent with the
continuum observed below 2 keV with the RGS. The X-ray flux varied by 20\%
during the 130~ks observation but no gross spectral variations were detected.
The spectrum does not require a reflection component, and shows no evidence for
absorption edges due to highly ionized Fe. The equivalent widths of the 6.4,
6.7 and 6.96 keV emission lines are $39^{+13}_{-12}$, $47^{+13}_{-13}$ and
$37^{+15}_{-15}$~eV respectively.  These equivalent widths are significantly
different to those normally observed in Seyfert galaxies, which typically have
stronger 6.4 keV lines and weaker 6.7 and 6.96 keV lines.  However, the ratios
of the three lines in M~81 are relatively similar to the those observed from
the Galactic Centre region.  The nucleus of M~81 does not appear to contain as
much cold, X--ray illuminated material as most Seyfert galaxies. The 6.7 and
6.96 keV emission lines might originate in photoionized gas within 0.1~pc of
the nuclear continuum source. Alternatively, these emission lines might have a
similar origin to the Galactic Centre Fe line emission, possibly through the
interaction of non-thermal cosmic ray electrons with the 0.2-0.6 keV thermal
plasma component of the M\,81 bulge.

\begin{acknowledgements}

We thank Michael Muno, Martin Weiskopf, Douglas Swartz and Alice Breeveld 
for helpful discussions and suggestions. 

\end{acknowledgements}

\end{document}